# Bright perovskite light-emitting electrochemical cell utilizing CNT sheets as a tunable charge injector

Masoud Alahbakhshi[a], Alexios Papadimitratos [d,e], Ross Haroldson[b], Aditya Mishra [c], Arthur Ishteev[f,g], Josef Velten [b,e], Qing Gu [a], Jason D. Slinker [b,e], and Anvar Zakhidov[*][b,c,d,e,f,g]

[a] Dept of Electr. Eng., [b]Physics Dept., [c]Dept of Mat. Sci. Eng., [d] NanoTech Institute, UT Dallas, TX, 75080; [e]Solarno Inc., Dallas, TX,75080; [f]NUST MISIS, Moscow, Russia; [g] ITMO University, St. Petersburg, Russia

## ABSTRACT

Organic-inorganic perovskite light-emitting devices have recently emerged as a reliable light source. Here, we developed a Single Layer Perovskite Light-Emitting Electrochemical Cells (SL-PeLEC) with laminated free-standing Carbon Nanotube Sheet (CNT) sheets as an effective charge electron injecting cathode electrode. The structure consists of bottom ITO-on-glass as a transparent electrode, the composite of $CsPbBr_3$:PEO:$LiPF_6$ with additive ionic salt as an emitting layer (EML) and 5 layers of CNT aerogel sheets as a top laminated cathode . Utilizing CNT free standing sheets laminated right on top of perovskite thin film in this simple single layer configuration has multiple benefits. Such CNT top cathode does not show any chemical degradation by reaction with halogens from perovskite, which is detrimental for metallic cathodes. Moreover, the formation of an internal p-i-n junction in perovskite EML composite layer by ionic migration under applied voltage bias and electric double layer (EDL) formation at each electrode interface is beneficially effecting CNT sheets by $Li^+$ ionic doping and raises their Fermi level, further enhancing electron injection. Besides, inspired by successes of ionic additives in LECs and electrochemical doping of perovskite with alkali metals, we leveraged a lithium salt, $LiPF_6$, within a $CsPbBr_3$:PEO composite matrix to achieve optimal ionic redistribution and doping effects in this SL-PeLEC. Although initially CNT electrode has slightly high sheet resistance, the SL-PeLEC device has a low turn-on voltage of 2.6v and a maximum luminance intensity of 530 cd/m$^2$, confirming the n-doping increased conductivity. This work provides a unique route toward flexible and bright perovskite LECs with stable and transparent CNT electrodes that can have injection efficiency tuned by poling induced ionic EDL-doping.

## 1. INTRODUCTION

Recently inorgano-organometallic halide perovskites materials with general formula $ABX_3$ (A is a smaller organic or inorganic cation, e.g., methylammonium ($CH_3NH_3^+$), formamidinium ($NH_2CHNH_2^+$), or cesium ion ($Cs^+$), B is a larger size divalent metallic cation ($Pb^{2+}$, $Sn^{2+}$), and X is halogen anion ($Cl^-$, $Br^-$, $I^-$)) have attracted much attention owing to unique optoelectronics and electronics properties such as low-temperature solution processability, high color purity with narrow spectral width (FWHM of 20 nm), bandgap tunability and large charge carrier mobility. [1] [2][3][4] To date, perovskite optoelectronic devices such as Lasers[5][6], solar cells [7], Photodetectors[8], light emitting diodes (LEDs)[9] and light emitting electrochemical cells[10][11] have shown promising performance in comparison with traditional inorganic devices. In this vein, perovskite light emitting electrochemical cells (PeLEC) have been invistigated as a reliable substitution of LEDs. These PeLEC devices leverage ion redistribution to achieve balanced and high carrier injection, resulting in high electroluminescence efficiency. Owing to this mechanism, LEC devices can be prepared from a simple architecture consisting of a single semiconducting composite layer sandwiched between two electrodes. In addition, they can operate at low voltages below the bandgap, yielding highly efficient devices.

Recently, many efforts have done to realize high performance PeLEC devices utilizing the benefits of ionic redistribution. Aygüler et al. utilized $FAPbBr_3$ nanoparticles with a trimethylolpropane ethoxylate polyelectrolyte and lithium triflate salt to reach luminance on the order of 1 cd m$^{−2}$ [12]. Zhang et al. proved that perovskite acts as a solid electrolyte and demonstrated light emission in both forward and reverse biases[13]. Also, Li et al. applied PEO inside the $MAPbBr_3$ to create a single layer device and reached up to 4000 cd m$^{−2}$ max luminance, with device performance indicative of both PeLEDs and polymer LECs[14]. Our group, recently demonstrated a significant step toward highly bright stable, efficient and very stable PeLEC with blending $CsPbBr_3$ with PEO, an electrolyte, and $LiPF_6$, an ionic additive. We successfully achieved a high luminance to 15000 cd/m$^2$ and long lifetimes extrapolating to 6700 h at 100 cd/m$^2$ [10] [11].

Figures 1a-c illustrate the distinct stages of PeLEC device operation with CNT sheet as an cathode electrode. Initially, ions are uniformly distributed in the film (Figure 1a). By applying a voltage, cations drift toward and accumulate near the cathode, and anions likewise move toward and pack near the anode (Figure 1b). This result in an electric double layer (EDL) formation at each electrode that induces higher electric fields, decreased width of the potential barriers, and enhanced injection of electrons and holes (Figure 1c)[15][16]. These injected carriers are transported through the bulk and radiatively recombine in the center of the device. In order to define a successful PeLEC operation, it needs to be considered some key features: 1) A sufficient concentration of mobile anions and cations in active layer; 2) Efficient transport of ions through the bulk for balanced EDL formation at the cathode and anode, leading to efficient electron and hole injection; 3) Facile transport of electrons and holes through the semiconductor (which, for our device, requires a percolating network of the perovskite); 4) Efficient light emission upon recombination of the electrons and holes in the bulk, typically supported by a high quantum yield of the film. A top view SEM image (Figure 1d) exhibits the dense network structure of highly porous CNT sheets on top of perovskite thin film.

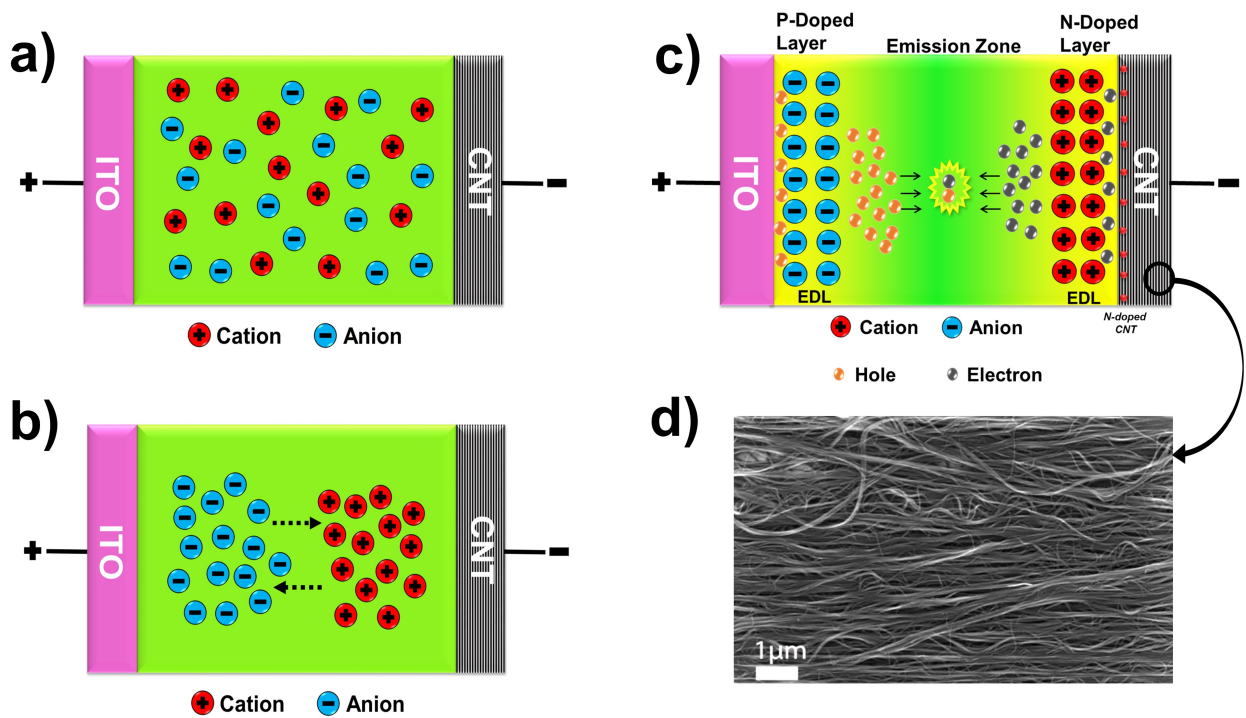

Figure 1. Mechanism of PeLEC device. a) Initial PeLEC state demonstrating ions uniformly distributed all over the active layer. b) Intermediate PeLEC state showing cations drifting toward the cathode and anions toward the anode. c) Steady-state PeLEC operation with ions accumulated at the electrodes and light emission upon current injection. d) Top views of the SEM images of CNT electrode

Even though LEC has numerous advantages such as uncomplicated fabrication process, variety of material to use as an active layer and in-situ formation of p-i-n junction layer, it might face some critical shortcomings that need to be addressed. One of the most significant disadvantages is the metal electrode. CNT as an ideal candidate for electrods has been vastly used due to high electrical conductivity, acceptable optical transparency and most importantly, chemical and thermal stability[17]. Yu et al reported an efficient and bright PeLED with a laminated CNT electrode[18]. Li et al applied a laminated CNT cathode to fabricate efficient perovskite solar cell[19]. Schulz et al showed electron donation from perovskite to the carbon nanotubes and a downward band bending toward the interface[20] Also, Bolink et al showed a flexible LEC employing single wall CNT anode[21]. In addition, we have introduced a new ambipolar perovskite light electrochemical cell configuration using single wall carbon nanotubes recently[22]. These results show that CNT is a promising candidate to apply as an electrode in optoelectronic devices.

In our system, we use an optimized ratio of $LiPF_6$ salt (0.5% weight ratio) which we have previously utilized to achieve high luminance and long lifetime PeLEC [10][11] into the $CsPbBr_3$:PEO (polyethylene oxide) composition. Here we present an easy and facile method to make a bright single layer PeLEC using multilayer carbon nanotube sheets. The PeLEC consists of a composite film of $CsPbBr_3$ composition deposited on top of ITO/glass substrate and 5 CNT sheets which deposited on top of perovskite composition thin film. We made 3 different devices (Device1: ITO/$CsPbBr_3$/CNT, Device2: ITO/$CsPbBr_3$:PEO/CNT and Device3: ITO/$CsPbBr_3$:PEO:$LiPF_6$/CNT) to compare the impact of $Li^+$ ions on the performance of perovskite LEC using CNT electrode. The device3 showed turn on voltage 2.6 v, reaching a brightness of 530 $cd/m^2$ with a current efficiency 0.3 cd/A at 6.5 v. The device2 turned on 3.3 v and a maximum brightness of 240 $cd/m^2$ at 6.5 v. In addition, we show that the polling effect can change the Fermi level of CNT and further enhance carrier injection. The formation of an internal p-i-n junction in perovskite emitting layer composite layer by ionic migration under applied voltage bias (polling) and electric double layer formation at each electrode interface is beneficially effecting CNT sheets by $Li^+$ ionic doping and raises their Fermi level. Our J-V results certainly demonstrated the $Li^+$ can penetrate deeper and effectively into the carbon nanotube structure by polling and further facilitate the charge injection into the emitting layer.

## 2. EXPERIMENTAL RESULTS AND DISCUSSIONS

The $CsPbBr_3$-based precursor solution was prepared by dissolving $PbBr_2$ and CsBr in a 1:1.5 molar ratio with PEO (10 mg/ml) in anhydrous DMSO solution. Then the perovskite precursor solution was stirred at 60 °C for dissolution overnight. When all solutions were dissolved, an empty vial was weighed and the desirable amount of PEO was added, then the weight difference before and after the PEO addition was measured to get an accurate weight of the viscous solution. The weight ratio of $CsPbBr_3$ to PEO was 100:80. Finally, these solutions were blended with 5 mg/ml DMSO solutions of $LiPF_6$ to prepare mixtures of 0.5% of lithium salt with the perovskite-polymer composition.

The ITO/glass substrates (resistance ~ 15 $\Omega$ $sq^{-1}$) were cleaned sequentially with detergent solution, deionized water, acetone, Toluene and 2-propanol in an ultra-sonication bath for 15 mins. Subsequently, the substrates were dried with nitrogen and treated for 20 min with UV-ozone. To obtain perovskite composition thin films, precursor solutions were spin-coated onto ITO substrates at 1200 rpm for 45 min. Then, all thin films were put under vacuum for 1 minute to have a uniform and pinhole-free thin film. Finally, all thin films were annealed at 150 °C for 15 seconds to remove the residual solvent. 5 CNT sheet layers were deposited on top of perovskite and then solidified by the HFS solvent. All measurements were conducted using a mechanical probe-station under a high vacuum < 10 mTorr. Current density-voltage (J-V) and luminance-voltage (L-V) characteristics were measured using a Keithley 236 source meter and a Photo Research PR-650 spectroradiometer increment. Additional current-only measurements were taken with a Keithley 2400 in a nitrogen glove box environment. The active area of the LEC is 9 $mm^2$.

As Figure 2a illustrates, all devices exhibited a green EL peak at 524 nm with an FWHM of 22 nm. Also, in figure 2b, the luminance vs voltage (L-V) of three devices is shown. The device3 reached to the maximum luminance about 530 $cd/m^2$ at 7v which is higher than device2 (240 $cd/m^2$ at 6.5 v) and device1 has 73 $cd/m^2$ at 8 v. The $CsPbBr_3$:PEO turned on at 3.3v but after adding $LiPF_6$ salt into the precursor, the device3 lighted up at 2.6v which denoting better formation of EDL and subsequently reducing the potential barrier width at each electrode active layer interface. The current density vs voltage (J-V) and Current Efficiency vs voltage (E-V) are shown in Figure3. The J-V curve for device2 demonstrated reaching to higher current density in comparison with device3. However, the higher current efficiency and brightness for device3 indicates in absence of lithium salt, unfilled trap states are present due to grain boundaries and vacancies that creates a nonradiative decay state in the middle of optical gap. Full performance metrics for all device formulations are provided in Table 1.

Table 1. Summarized performance of our PeLEC devices with CNT electrode.

| Devices | Turn-on Voltage (v) | Max Current Efficiency (cd/A) | Max Luminance (cd/m$^2$) | Max Current Density (mA/cm$^2$) |
|---|---|---|---|---|
| ITO/CsPbBr$_3$/CNT | 6.2 | 0.06 | 73 @ 8v | 150 |
| ITO/CsPbBr$_3$:PEO/CNT | 3.3 | 0.1 | 240 @ 6.5v | 335 |
| ITO/CsPbBr$_3$:PEO:LiPF$_6$/CNT | 2.6 | 0.3 | 530 @ 7v | 220 |

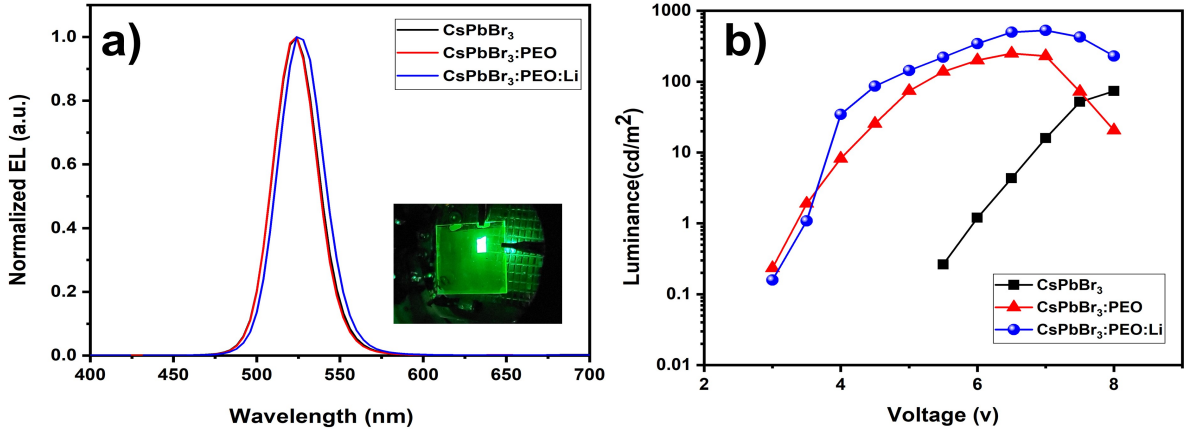

Figure 2. a) Electroluminescence spectra (EL) and b) Luminance vs voltage (L-V) of three PeLEC devices with CNT as a cathode electrode. The inset in a) shows the EL of device3 operating at 6.5v.

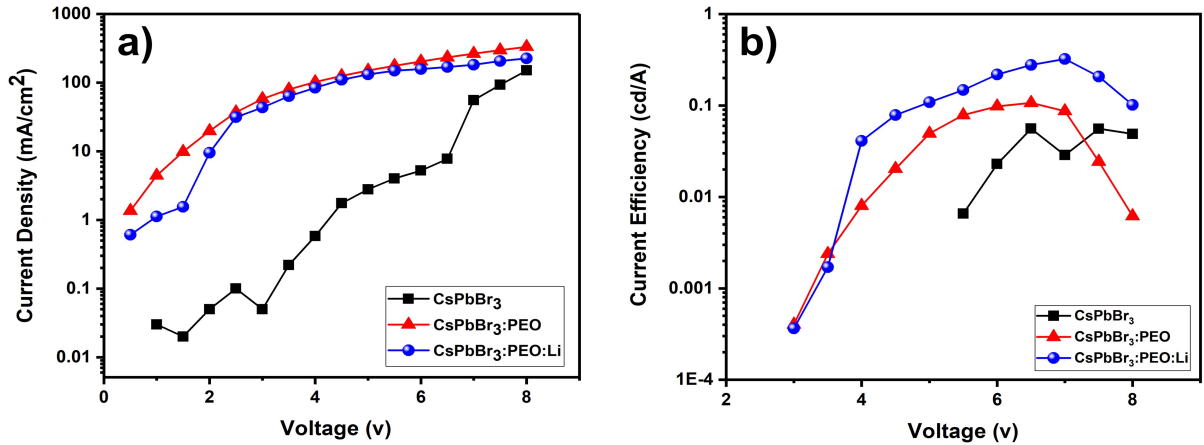

Figure 3. a) Current Density vs voltage (J-V) and b) Current Efficiency vs voltage (E-V) of three PeLEC device with CNT as a catode electrode.

To shed light on the CNT role in PeLEC performance and ions migrations, we successfully observed that pre-voltage bias in a specific time (polling) can tune the intrinsic properties of carbon nanotube structure. This means the perovskite composition structure under constant voltage slightly higher than turn-on voltage, electrical polling form more accumulation of negative and positive ions at the electrode interfaces, then formation of EDL and consequently raise the Fermi level of CNT electrode. This upward shifting of the Fermi level feasibly enhances and facilities the injection of electrons through the carbon structure into the perovskite structure.

Figure 4a shows the current density (CD) changing of device2 under applied voltage at 3.5v. The initial CD without any polling is around 47 mA/cm2. After keeping the constant voltage for 2 mins, we immediately switched to the forward sweep from -4 to 8 v. The results demonstrated that not only we could enhancing the CD to 150 mA/cm$^2$ after 2 mins polling, but also the luminance increased significantly. This evidently shows an upward shift of Fermi level of CNT after polling which is correlated to positive cations like $Cs^+$ and negative ions like $Br^-$ moving toward the perovskite/CNT and the perovskite/ITO interfaces, respectively. The aggregated cations ($Cs^+$) would penetrate through the perovskite/CNT interface and fill the space between carbon nanotube bundles due to small ionic radius of cations. Such adsorption of positive ions inside of CNT pores correlates well with the n-type doping of CNT and leads to the higher Fermi level of CNT electrode and eventually more electron injection. It is worth mentioning that the longer polling is applied, the more ions can accumulate at the interfaces, as EDL which in turn may create more some degradation at the interfaces. This might be the reason for lower enhancement of CD after 10 mins polling.

Figure 4b demonstrates CD changing of device3 under constant voltage at 3 v. In contrary to device2, in presence of the Li ions, the current density keeps increasing even after 10 mins polling. In fact, the positive $Li^+$ cations in the $CsPbBr_3$:PEO:$LiPF_6$ composition, due to having lower activation energy in comparison with $Cs^+$ and its vacancies $V^+$ as well as the smaller ionic radius, can paly important role to create faster and higher charge density EDL formation as we proved in our latest work[11]. This results in a stronger n-doped region near to the CNT cathode and consequently more penetration of $Li^+$ into the carbon nanotube sheet porous structure. In the same vein, the $PF_6^-$ counterion dissociated from $Li^+$ by the PEO matrix can be accumulated at the interface of ITO and make a strong EDL region. Plus, we can observe that polling voltage can significantly reduce the hysteresis of CD curve because of faster ionic redistribution which is comparable with our previous work[10]. These results show that in the presence of lithium additive, the n-doped CNT electrode can remarkably increase the injection of electrons due to the raising of the Fermi level. Figure 4c illustrates the band energy diagram of device2 and device3 as well as the $Cs^+$ and $Li^+$ penetration into the CNT sheets network.

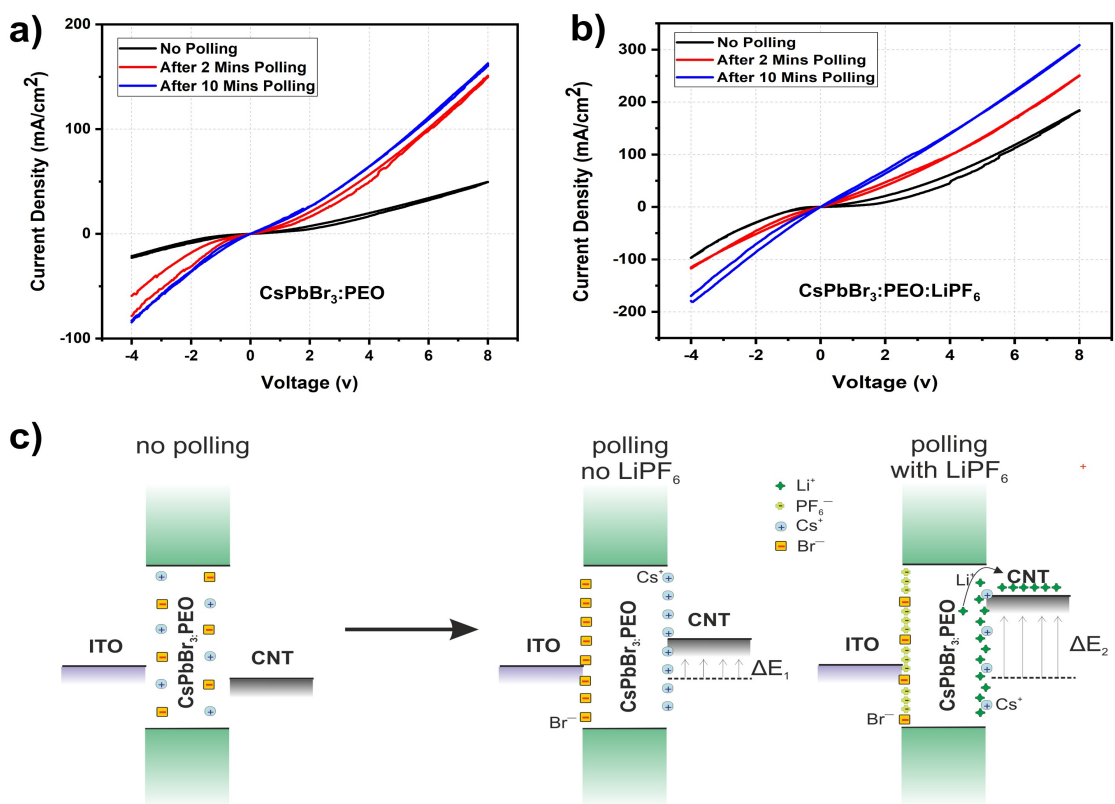

Figure 4. The polling effect on a) device2 under constant voltage of 3.5v and b) device3 with added optimal $LiPF_6$ under constant voltage of 3v. c) the schematics of the band energy diagram, showing the raise of Fermi level of CNT upon $Cs^+$ penetration into CNT sheets network for device2 as well as $Cs^+$ and $Li^+$ penetration for device3.

## 3. CONCLUSION

We reported stable CNT electron injectors in single layer perovskite light emitting electrochemical cells (PeLEC) based on both pristine $CsPbBr_3$:PEO, and same device with an optimal concentration of $LiPF_6$ ionic additives, which exhibited bright stable light emission. It is demonstrated that poling effects cause the n-doping of CNT electrodes via the EDL formation by positive ions penetration from the perovskite emitting layer into CNT porous network. The increase of the current in J-V curves is studied in both types of PeLEC (with and without Li ionic additives) upon poling and it is clearly more pronounced in $Li^+$ added case, due to better n-doping via faster and deeper penetration of small mobile $Li^+$ ions as compared to $Cs^+$ ions. Subsequently, the enhanced electron injection from n-doped CNT takes place from a higher raised Fermi level in the case of $Li^+$ ions leading to brighter Electroluminescence. The combination of stable (and possibly transparent) doped carbon nanotube electrodes via poling induced p-i-n formation further enhanced by n-doping of CNT allows stable electroluminescence that has multiple advantages as compared to conventional metallic cathodes. It should be mentioned that in full analogy with described here n-doping, the p-type doping of multi wall CNT can be induced by opposite polarity of poling by penetration of negative ions ($Br^-$, $PF_6^-$., etc.) and thus leads to ambipolar PeLEC [22] that can operate in AC regime, as we will describe in forthcoming paper.

## ACKNOWLEDGEMENTS

A.Z. acknowledges support from the Welch Foundation (AT-1617) and from the Ministry of Education and Science of the Russian Federation (14.Y26.31.0010). Q.G. acknowledges support from the Welch Foundation (AT-1992-20190330). J.D.S. acknowledges support from the National Science Foundation (ECCS 1906505). A.I. and A.Z also acknowledge financial support of Increase Competitiveness Program of NUST "MISiS" (No. K2-2017-007).